\documentclass[prd,aps,a4paper,nofootinbib,twocolumn]{revtex4}  

\newif\ifusesec
\usesectrue  
   
\usepackage{graphicx} 
\usepackage{mathrsfs}
\usepackage{amsmath,amsfonts,amssymb}
\usepackage{multirow}

\newcommand{\beq}{\begin{equation}}
\newcommand{\eeq}{\end{equation}}

\begin{document}

\title{Gauge-fixing for the completion problem of reconstructed metric perturbations of a Kerr spacetime}

\author{Donato \surname{Bini}}
\author{Andrea \surname{Geralico}}

\affiliation{Istituto per le Applicazioni del Calcolo ``M. Picone,'' CNR, I-00185 Rome, Italy}

\date{\today}

\begin{abstract}
We provide a prescription to solve the metric completion problem in gravitational self-force calculations on a Kerr spacetime by fixing the remaining gauge freedom.
We discuss the explicit example of eccentric equatorial orbits, recovering all limiting cases already studied in the literature of eccentric orbits in  Schwarzschild as well as circular orbits in both Schwarzschild and Kerr spacetimes. 
\end{abstract}

\maketitle

\section{Introduction}

The issue of metric completion in gravitational perturbation theory is a longstanding problem, since the seminal works of Regge, Wheeler and Zerilli (RWZ) \cite{Regge:1957td,Zerilli:1971wd}, who provide the necessary formalism to study the first order perturbations of the Schwarzschild solution. This approach was soon applied to the analysis of gravitational radiation emitted by a small compact body \cite{Davis:1971gg}, which captured the main interest of the gravity community. Since the computation of energy and angular momentum fluxes only involves the radiating part of the metric, the problem of low multipoles remained overlooked for many years.
Zerilli himself showed that these multipoles simply correspond to \lq\lq shifts'' in the mass and angular momentum of the background spacetime. Their relevance for self-force calculations became clear after the work of Detweiler and Poisson \cite{Detweiler:2003ci}, who emphasized that  such shifts are as important as the radiating multipoles for describing the motion of a small body orbiting a black hole, since their contribution to the (conservative piece of the) self-force affects the dynamics even at the Newtonian level.

The nonradiative part of the perturbed metric thus plays a crucial role in gravitational self-force (GSF) calculations, for which the complete reconstruction of the metric perturbations is necessary to compute the various orbital invariants, like redshift, periastron advance, spin-precession angle and tidal invariants, which encode a gauge-invariant information on the dynamics, useful to compare results from different approximation methods, either analytic or numeric.
We refer to Ref. \cite{Barack:2018yvs} for a recent review on GSF computational techniques, and on the increasing importance of GSF calculations for an even more accurate modelling of the dynamics of extreme-mass-ratio inspirals (EMRIs), which will be a primary source of gravitational waves for the planned low-frequency space-based detector eLISA \cite{elisa}.
The RWZ formalism provides all necessary tools for fully determining the nonradiative piece of the perturbed metric in the Schwarzschild case, directly using the Zerilli's results together with suitable gauge adjustments to ensure regularity as well as asymptotic flatness of the perturbation \cite{Detweiler:2003ci,Bini:2014ica,Hopper:2015icj}.
Solving the same problem in Kerr, instead, is still a challenge since many years.

The basic theory of gravitational perturbations of a Kerr spacetime was developed by Teukolsky using the Newman-Penrose formalism \cite{Teukolsky:1973ha,Press:1973zz,Teukolsky:1974yv}. The Einstein field equations combined with the Bianchi identities lead to a single master wave equation, the Teukolsky equation, for the perturbed Weyl scalars $\psi_0$ ($s=2$) or $\psi_4$ ($s=-2$), which can be solved by separation of variables (using spheroidal harmonics and Fourier decomposition).
The radiative part of the metric perturbation can then be reconstructed from a scalar function, the Hertz potential, in a radiation gauge through the Chrzanowski-Cohen-Kegeles (CCK) procedure \cite{Cohen:1974cm,Chrzanowski:1975wv,Kegeles:1979an}.
The latter was originally developed for vacuum perturbations, and more recently extended to metric perturbations sourced by a particle moving along a bound geodesic orbit around a Kerr black hole \cite{Ori:2002uv}. Despite the appearance of irregular behaviors (string-like singularities) caused by the presence of matter (both within and outside the region where the source is located) \cite{Pound:2013faa}, the use of radiation gauge metric perturbations and related CCK formalism to obtain self-force corrections to the particle's motion is now well established in GSF theory \cite{Keidl:2010pm,Shah:2010bi,Shah:2012gu,vandeMeent:2015lxa}.  

The perturbed metric should then be completed by the nonradiative modes, which cannot be determined by the Teukolsky equation, since the spheroidal harmonics are not defined for $l<|s|=2$, in contrast with the Schwarzschild case, where these lower multipoles associated with $l=0,1$ can be expressed appropriately in terms of spherical harmonics using the RWZ formalism.
The remaining part must be stationary and axially symmetric, simply corresponding to mass and angular momentum perturbations of the Kerr background in the vacuum region away from the particle's location, up to gauge modes \cite{Wald:1973,Wald:1978vm}.
This completion piece can be computed, in principle, in any gauge, and no general prescription has been found yet.

According to Ref. \cite{Pound:2013faa} a regular (\lq\lq no-string'') solution can be formed by joining together the regular sides of two \lq\lq half-string'' solutions along a  hypersurface containing the particle's world line, supporting a gauge discontinuity (and possibly distributional singularities).
The latter separates the spacetime region spatially inside the particle's location (interior region, $-$) from that outside it (exterior region, $+$), so that the full metric perturbation can be split in three different contributions \cite{Barack:2017oir}
\beq
\label{hdef}
h_{\alpha\beta}^{\pm}=h_{\alpha\beta}^{{\rm rec}\pm}+h_{\alpha\beta}^{{\rm comp}\pm}+h_{\alpha\beta}^{{\rm gauge}\pm}\,.
\eeq
The reconstructed metric perturbation $h_{\alpha\beta}^{{\rm rec}\pm}$ is obtained by the CCK procedure, and represents the radiative part also referred to as $h_{\alpha\beta}^{{\rm rad}\pm}$ below.
The sum of the completion piece and the gauge piece instead gives the nonradiative part $h_{\alpha\beta}^{{\rm nonrad}\pm}$.
A method to compute $h_{\alpha\beta}^{{\rm comp}\pm}$ has been recently proposed in Ref. \cite{Merlin:2016boc} for eccentric equatorial motion (see also references therein for a review of previous attempts), based on the construction of certain gauge-invariant fields from the full perturbed metric. Imposing the continuity of these quantities across the hypersurface containing the particle world line fixes the completion piece of the metric perturbation in a way that in the spacetime region outside such a hypersurface the mass and angular momentum are given by the particle's conserved energy and angular momentum, whereas both vanish in the region inside it ($h_{\alpha\beta}^{{\rm comp}-}\equiv0$).
This result has been then generalized to any bound orbit around a Kerr black hole in Ref. \cite{vandeMeent:2017fqk}. 

A last problem still remains unsolved, how to determine the gauge part $h_{\alpha\beta}^{{\rm gauge}\pm}$.
The \lq\lq gauge-smoothing'' of the perturbation across the particle's world line is crucial for the GSF calculation of orbital invariants.
In fact, if the gauge part is not calibrated properly, the computation will not give the correct result.
The orbital invariants are indeed invariants under a class of gauge transformations which are sufficiently smooth functions of the coordinates and also preserve the symmetries of the perturbed spacetime, e.g., the helical symmetry for circular orbits \cite{Detweiler:2008ft}. 
In contrast, the perturbed metric is discontinuous at the location of the particle, so that a gauge transformation connecting the interior and exterior parts must be discontinuous too. This fact may affect or not the invariance of the quantity constructed with the full perturbation.
A useful check is the comparison with the results obtained through different methods, e.g., within the PN theory.
In any case, one should require the gauge part to share the same regularity as well as symmetry properties as the full perturbation.
For instance, in the simplest case of circular motion one can reduce the gauge freedom by demanding that the metric perturbation preserves the helical symmetry, besides the usual conditions of regularity and asymptotic flatness.
Further imposing the continuity of certain metric components has been shown in Ref. \cite{Shah:2015nva} to completely fix the gauge on a Schwarzschild background. 
Unfortunately, the same reasoning cannot be applied to more general situations, like eccentric orbits, and still for circular orbits in Kerr.
A different strategy has been suggested in Refs. \cite{vandeMeent:2015lxa,vandeMeent:2016hel,Bini:2018ylh}, but not fully implemented yet, consisting in requiring the continuity at the particle position of suitably chosen \lq\lq quasi-invariant'' fields built with the metric perturbation (see also the related discussion in Section 7.6 of Ref. \cite{Barack:2018yvs}).
So far, the only possibility to overcome this difficulty has been to choose a \lq\lq reasonable'' gauge, do the calculation of an orbital invariant (or any other gauge-invariant function), and check the agreement of the first few terms of its PN expansion with available PN results, eventually adjusting the gauge part if needed.
This is the reason why the redshift invariant was the first orbital invariant to be computed in a Kerr spacetime \cite{Shah:2012gu,Bini:2015xua,Kavanagh:2015lva,vandeMeent:2015lxa,Bini:2016dvs,Bini:2019lcd}.
In fact, it is defined in terms of the double contraction of the perturbed metric with the particle's four-velocity, which is a continuous function across the particle already at the level of individual radiative multipole modes, before summing over them and regularizing.
Therefore, the knowledge of the nonradiative part in the exterior region (where the gauge part must vanish for regularity reason, as we will show below) suffices in the case of the redshift.
As soon as one is considering more singular quantities at the particle's location, like gyroscope precession (i.e., a connection term) and tidal invariants (i.e., curvature terms), involving first and second derivatives of the perturbed metric, respectively, one cannot expect such a simple feature, so that a more general procedure is needed. 

We provide here a general prescription to completely fix the remaining gauge freedom, with specific applications to eccentric equatorial orbits in the Kerr spacetime.  
The exterior part vanishes identically ($h_{\alpha\beta}^{{\rm gauge}+}\equiv0$) due to the request of asymptotic flatness.
The components of the most general gauge vector generating the interior part $h_{\alpha\beta}^{{\rm gauge}-}$, instead, are determined by requiring that the causality condition of the particle's four velocity be preserved at every spacetime point, including the location of the particle, and by imposing on the full perturbed metric the Ricci identity across the hypersurface containing the particle's world line.

The paper is organized as follows. 
We will start by reviewing the problem of low multipoles in the case of a Schwarzschild spacetime and its solution in the RWZ gauge. In the simplest case of perturbations due to a particle moving along an equatorial circular orbit we will show that the request of asymptotic flatness and continuity of certain metric components implies a gauge-adjustment of the nonradiative part of the metric by a necessarily discontinuous gauge vector, which generates an additional energy momentum tensor contributing to the Einstein's equations. 
We will pass then to the Kerr case, discussing the various parts of the perturbations in the same line of reasoning: radiative and nonradiative, the latter further splitting into completion and gauge parts.
Finally, we will provide a prescription to completely determine the gauge part, and apply it to the case of eccentric equatorial orbits. 
We will use geometrical units $G=1=c$.

\section{Perturbations on a Schwarzschild spacetime}

Let us start by reviewing the gauge problem for gravitational perturbations on a Schwarzschild black hole background, with line element written in standard spherical-like coordinates $(t,r,\theta,\phi)$ given by
\beq 
\label{metric}
ds^2 = -N^2d t^2 + N^{-2} d r^2 
+ r^2 (d \theta^2 +\sin^2 \theta d \phi^2)\,,
\eeq
where $N=(1-2M/r)^{1/2}$ denotes the lapse function.
The nonradiative part of the metric is obtained by solving the perturbation equations corresponding to the lowest multipoles $l=0,1$ in a spherical harmonic decomposition of the metric, following the original approach of Zerilli.
We will consider below the simplest case of a particle moving along a circular equatorial orbit of radius $r=r_0$, with four velocity 
\beq
U=U^t(\partial_t +\Omega \partial_\phi)\,,
\eeq
with
\beq
U^t=\left(1-\frac{3M}{r_0}\right)^{-1/2}\,,\qquad 
\Omega=\sqrt{\frac{M}{r_0^3}}\,.
\eeq
The particle's energy-momentum tensor
\begin{eqnarray}
T^{\mu\nu}&=&\mu\frac{U^\mu U^\nu}{r_0^2 U^t}\delta(r-r_0)\delta\left(\theta-\frac{\pi}{2}\right)\delta(\phi-\Omega t)
\nonumber\\
&=&\mu\sum_{lm}\frac{U^\mu U^\nu}{r_0^2 U^t}\delta(r-r_0)Y_{lm}(\theta,\phi)Y^*_{lm}\left(\frac{\pi}{2},\Omega t\right)
\,,
\end{eqnarray}
can be decomposed into a radiative part 
\beq
T_{\mu\nu}^{\rm rad}=\sum_{l\geq2,m}T_{\mu\nu}^{lm}\,,
\eeq
and a completion part 
\beq
T_{\mu\nu}^{\rm comp}=\sum_{l=[0,1],m}T_{\mu\nu}^{lm}\,,
\eeq
associated with the radiative part $h_{\alpha\beta}^{\rm rad}$ ($l\geq2$) and the completion part $h_{\alpha\beta}^{\rm comp}$ ($l=0,1$) of the perturbation, respectively.
We will refer to the original work of Zerilli \cite{Zerilli:1971wd} for notation and conventions.

After decomposing both the perturbed metric and the energy-momentum tensor in spherical harmonics, the Einstein's field equations reduce to two sets of perturbation equations of different parity: the even sector consists of seven equations for the perturbation functions $H_0$, $H_1$, $H_2$, $K$, $h_0^{\rm even}$, $h_1^{\rm even}$, whereas the odd sector consists of three equations for the perturbation functions $h_0^{\rm odd}$, $h_1^{\rm odd}$, $h_2^{\rm odd}$.
This general form of the perturbation can be simplified by performing a suitable gauge choice.
Consider the infinitesimal coordinate transformation 
\beq
x{}'^\mu=x^\mu+\xi^\mu\,,
\eeq
where the infinitesimal displacement $\xi^\mu$ is a function of coordinates $x^\mu$ and transforms like a vector. 
This transformation induces the following transformation of the metric perturbation 
\beq
h_{\mu\nu}'=h_{\mu\nu}-2\xi_{(\mu;\nu)}\,,
\eeq
where the multipole expansion of the second term can be easily obtained by expanding the gauge vector $\xi$ in vector harmonics of both types, i.e., $\xi=\xi^{\rm even}+\xi^{\rm odd}$, with 
\begin{eqnarray}
\xi^{\rm even}&=&\sum_{lm}\left[A_0Y_{lm}dt+A_1Y_{lm}dr\right.\nonumber\\
&&\left.
+A_2\left(\frac{\partial Y_{lm}}{\partial \theta}d\theta+\frac{\partial Y_{lm}}{\partial \phi}d\phi\right)\right]\,,
\end{eqnarray}
and
\beq
\xi^{\rm odd}=\sum_{lm}\frac{A_3}{\sin\theta}\left(\frac{\partial Y_{lm}}{\partial \phi}d\theta-\sin^2\theta\frac{\partial Y_{lm}}{\partial \theta}d\phi\right)\,,
\eeq
where $A_\alpha=A_{\alpha\,lm}(t,r)$.
Under such a transformation the even and odd perturbation functions change according to
\begin{eqnarray}
H_0'&=&H_0-\frac{2}{N^2}\partial_t A_0+\frac{1-N^2}{r}A_1\,,\nonumber\\
H_1'&=&H_1-\partial_t A_1-\partial_r A_0+\frac{1-N^2}{rN^2}A_0\,,\nonumber\\
H_2'&=&H_2-2N^2\partial_r A_1-\frac{1-N^2}{r}A_1\,,\nonumber\\
K'&=&K-\frac{2N^2}{r}A_1+\frac{2}{r^2}A_2\,,\nonumber\\
G'&=&G-\frac{2}{r^2}A_2\,,\nonumber\\
h_0'^{\rm even}&=&h_0^{\rm even}-A_0-\partial_t A_2\,,\nonumber\\
h_1'^{\rm even}&=&h_1^{\rm even}-A_1+\frac{2}{r}A_2-\partial_r A_2\,,
\end{eqnarray}
and 
\begin{eqnarray}
h_0'^{\rm odd}&=&h_0^{\rm odd}+\partial_t A_3\,,\nonumber\\
h_1'^{\rm odd}&=&h_1^{\rm odd}-\frac{2}{r}A_3+\partial_r A_3\,,\nonumber\\
h_2'^{\rm odd}&=&h_2^{\rm odd}-2A_3\,,
\end{eqnarray}
respectively.
The Regge-Wheeler gauge sets to zero the functions $G$, $h_0^{\rm even}$, $h_1^{\rm even}$ and $h_2^{\rm odd}$.

If the gauge vector is a sufficiently smooth function of the coordinates, i.e., at least twice differentiable, then the gauge part $h_{\mu\nu}^{\rm gauge}=2\xi_{(\mu;\nu)}$ is a \lq\lq pure gauge'' metric perturbation, which is automatically a solution of the vacuum Einstein's equations $\delta G_{\mu\nu}[h_{\alpha\beta}^{\rm gauge}]=0$.
In that case, the radiative part of the perturbation satisfies the perturbed equations 
\beq
\delta G_{\mu\nu}[h_{\alpha\beta}^{\rm rad}]=8\pi T_{\mu\nu}^{\rm rad}\,,
\eeq
whereas the nonradiative one 
\beq
\label{EFEnonradschw}
\delta G_{\mu\nu}[h_{\alpha\beta}^{\rm comp}]=8\pi T_{\mu\nu}^{\rm comp}\,,
\eeq
so that 
\begin{eqnarray}
\delta G_{\mu\nu}[h_{\alpha\beta}^{\rm rad}]+\delta G_{\mu\nu}[h_{\alpha\beta}^{\rm comp}]&=&8\pi (T_{\mu\nu}^{\rm rad}+T_{\mu\nu}^{\rm comp})\nonumber\\
&=& 8\pi T_{\mu\nu}\,.
\end{eqnarray}
In contrast, we will show below that in order to get a nonradiative metric perturbation with the desired properties the (nonradiative part of the) gauge vector must be a discontinuous function across the particle's world line, thus generating an Einstein tensor or equivalently an additional energy-momentum tensor with distributional singularities there to be added to the right-hand-side of Eq. \eqref{EFEnonradschw}.

\subsection{The monopole mode $l=0$}

The $l=0$ mode represents the perturbation in the total mass of the system due to the addition of the particle's conserved Killing energy $\delta M\equiv E=-U_t$. 
For this mode there are two gauge degrees of freedom and the choice done by Zerilli is
\beq
H_1^Z(t,r)=0=K^Z(t,r)\,.
\eeq
The solution for the remaining perturbation functions $H_0$ and $H_2$ is
\beq
H_0^Z=H_2^Z=\frac{a}{r-2M}\theta (r-r_0)\,,
\eeq
with
\beq
a=2\sqrt{4\pi}\,\delta M\,,
\eeq
and $\theta(x)$ denoting the Heaviside step function.
The nonvanishing metric components are then given by
\begin{eqnarray}
h_{tt}^Z&=&\frac{N^2H_0}{\sqrt{4\pi}}=\frac{2\delta M}{r}\theta (r-r_0)
\,,\nonumber\\
h_{rr}^Z&=&\frac{H_2}{\sqrt{4\pi}N^2}=\frac{2\delta M}{rN^4}\theta (r-r_0)\,.
\end{eqnarray}

One can use the gauge freedom to make the $h_{tt}$ component continuous across the particle's position without modifying the $h_{rr}$ component. 
This is done by choosing 
\beq
A_0=-\frac{at}{2r}\frac{r-2M}{r_0-2M}\theta (r_0-r)\,,\qquad
A_1=0\,,
\eeq
so that
\begin{eqnarray}
H_0&=&\frac{a}{r_0-2M}\theta (r_0-r)+\frac{a}{r-2M}\theta (r-r_0)\,,\nonumber\\
H_1&=&-\frac{at}{2r_0}\delta (r-r_0)\,,\nonumber\\
H_2&=&\frac{a}{r-2M}\theta (r-r_0)\,.
\end{eqnarray}
The new metric perturbation is then given by
\begin{eqnarray}
h_{tt}'&=&\frac{2\delta M}{r}\left[
\frac{rN^2}{r_0N_0^2}\theta (r_0-r)+\theta (r-r_0)
\right]\,,\nonumber\\
h_{tr}'&=&-\frac{2\delta Mt}{2r_0}\delta (r-r_0)
\,,\nonumber\\
h_{rr}'&=&\frac{2\delta M}{rN^4}\theta (r-r_0)\,.
\end{eqnarray}

\subsection{The dipole $l=1$ odd mode}

The $l=1$ odd mode represents the angular momentum perturbation $\delta J\equiv L=U_\phi$ added by the particle to the system. 
Zerilli takes $h_1^{{\rm odd}\,Z}=0$, and the remaining perturbation equations imply
\beq
h_0^{{\rm odd}\,Z}=b\frac{r_0}{r}\theta (r-r_0)\,,
\eeq
with
\beq
b=2\sqrt{\frac{4\pi}{3}}\frac{\delta J}{r_0}\,,
\eeq
and only the $m=0$ mode is nonzero.
The only nonvanishing metric component is then given by
\begin{eqnarray}
h_{t\phi}^Z&=&-\sqrt{\frac{3}{4\pi}}h_0^{{\rm(odd)}\,Z}\sin^2\theta\nonumber\\
&=&-\frac{2\delta J}{r}\theta (r-r_0)\sin^2\theta\,.
\end{eqnarray}

Choosing 
\beq
A_3=bt\frac{r^2}{r_0^2}\theta (r_0-r)\,,
\eeq
allows one to make the $h_{t\phi}$ component continuous across the particle's position.
In fact, the new function $h_0^{\rm odd}$ becomes
\beq
h_0'^{\rm odd}=b\left[\frac{r^2}{r_0^2}\theta (r_0-r)+\frac{r_0}{r}\theta (r-r_0)\right]\,,
\eeq
but a nonzero function $h_1^{\rm odd}$ is also generated
\beq
h_1'^{\rm odd}=-bt\delta(r-r_0)\,.
\eeq
The new metric perturbation is then given by
\begin{eqnarray}
h_{t\phi}'&=&-\frac{2\delta J}{r}\left[\frac{r^3}{r_0^3}\theta (r_0-r)+\theta (r-r_0)\right]\sin^2\theta\,,\nonumber\\
h_{r\phi}'&=&-\sqrt{\frac{3}{4\pi}}h_1'^{{\rm(odd)}}\sin^2\theta\nonumber\\
&=&\frac{2\delta J}{r_0}t\delta(r-r_0)\sin^2\theta\,.
\end{eqnarray}

\subsection{The dipole $l=1$ even mode}

The $l=1$ even mode is related to the shift of the center of momentum of the system. 
Zerilli assumes $K^Z=0$ in addition to $h_0^{{\rm even}\,Z}=0=h_1^{{\rm even}\,Z}$.
The solution for the remaining metric functions in the Zerilli gauge turns out to be
\begin{eqnarray}
H_0^{Z}&=&
\frac{c(t)}{3}\frac{r_0N_0^2}{r^2N^4}\left(1-\Omega^2\frac{r^3}{M}\right)\delta M\theta (r-r_0)\,,\nonumber\\
H_1^{Z}&=&
imc(t)r\Omega\frac{r_0N_0^2}{r^2N^4}\delta M\theta (r-r_0)
\,,\nonumber\\
H_2^{Z}&=&
c(t)\frac{r_0N_0^2}{r^2N^4}\delta M\theta (r-r_0)\,,
\end{eqnarray}
with
\beq
c(t)=-2\sqrt{6\pi}m e^{-im\Omega t}\,.
\eeq
The nonvanishing metric components are then given by
\begin{eqnarray}
h_{tt}^{Z}&=&
2\frac{r_0N_0^2}{r^2N^2}\left(1-\Omega^2\frac{r^3}{M}\right)\delta M\theta (r-r_0)\sin\theta\cos\bar\phi\,,\nonumber\\
h_{tr}^{Z}&=&
-6r\Omega\frac{r_0N_0^2}{r^2N^4}\delta M\theta (r-r_0)
\sin\theta\sin\bar\phi\,,\nonumber\\
h_{rr}^{Z}&=&
6\frac{r_0N_0^2}{r^2N^6}\delta M\theta (r-r_0)
\sin\theta\cos\bar\phi\,,
\end{eqnarray}
where $\bar\phi=\phi-\Omega t$.
Therefore, the solution is not asymptotically flat.
In vacuum, a dipole even-parity perturbation can be completely removed by a gauge transformation, as shown by Zerilli.
This is no more true in the nonvacuum case, when a source term is present given by the the particle's energy momentum tensor.
However, one can always perform a \lq\lq singular gauge'' transformation \cite{Zerilli:1971wd,Detweiler:2003ci} which removes the perturbation in the vacuum region outside of the particle's location, leaving a nonvanishing contribution just at $r=r_0$.  

Let $h_0'^{\rm even}=0=h_1'^{\rm even}$, in order that the RW gauge be still holding. This implies
\beq
A_0=-\partial_t A_2\,,\qquad
A_1=\frac{2}{r}A_2-\partial_r A_2\,.
\eeq
Choosing then
\beq
A_2=f_2(t)\frac{r^2}{r-2M}\theta (r-r_0)\,,\quad
f_2(t)=-\frac{c(t)}{6}\frac{r_0}{M}N_0^2\delta M\,,
\eeq
leads to the new metric components of the type
\beq
h'_{\alpha\beta}=h'_{\alpha\beta}{}^\delta \delta(r-r_0)+h'_{\alpha\beta}{}^{\delta'} \delta'(r-r_0)\,,
\eeq
that is, explicitly
\begin{eqnarray}
h_{tt}'&=&
2N_0^2\delta M\delta(r-r_0)\sin\theta\cos\bar\phi\,,\nonumber\\
h_{tr}'&=&
-\frac{2}{M}r_0^2\Omega\delta M\delta(r-r_0)\sin\theta\sin\bar\phi\,,\nonumber\\
h_{rr}'&=&
\frac{2\delta M}{M}\frac1{N_0^2}[
(r_0-M)\delta(r-r_0)\nonumber\\
&&
-r_0^2N_0^2\delta'(r-r_0)]
\sin\theta\cos\bar\phi\,,\nonumber\\
h_{\theta\theta}'&=&
-\frac{2}{M}r_0^3N_0^2\delta M\delta(r-r_0)\sin\theta\cos\bar\phi\,,\nonumber\\
h_{\phi\phi}'&=&
\sin^2\theta h_{\theta\theta}'\,,
\end{eqnarray}
which are all delta-singular (meaning that they generically include $\delta(r-r_0)$, $\delta'(r-r_0)$, etc. contributions) at the particle's position.

\subsection{Final form of the nonradiative part of the perturbation}

In the Zerilli gauge the completion part of the metric perturbation can be written as
\beq
\label{hcompschw}
h_{\alpha\beta}^{\rm comp}\equiv h_{\alpha\beta}^{Z}
=\tilde h_{\alpha\beta}^{Z}\theta(r-r_0)\,,
\eeq
with nonvanishing components
\begin{eqnarray}
\tilde h_{tt}^{Z}&=&\frac{2\delta M}{r}
\,,\quad
\tilde h_{rr}^{Z}=\frac{2\delta M}{rN^4}
\,,\nonumber\\
\tilde h_{t\phi}^{Z}&=&-\frac{2\delta J}{r}\sin^2\theta\,.
\end{eqnarray}
Therefore, the interior metric is identically vanishing, and the monopole and dipole (exterior) perturbations describe the sudden shift in mass and angular momentum parameters induced by the particle occurring at $r=r_0$, respectively, modulo the presence of gauge terms associated with a change in the system's center of mass, which are delta-singular there \cite{Zerilli:1971wd,Detweiler:2003ci}. The latter will be included in the gauge piece.

The gauge freedom can be used to modify the final form of the perturbation by requiring, e.g., asymptotic flatness and the continuity of certain metric components. 
The gauge part of the perturbation
\beq
h_{\alpha\beta}^{\rm gauge}=2\xi_{\,(\alpha; \beta)}\,,
\eeq
is generated by the gauge field
\begin{eqnarray}
\label{xischwcirc}
\xi&=&-\frac{A_0}{N^2}Y_{00}\partial_t-\frac{A_3}{r^2\sin\theta}\frac{\partial Y_{10}}{\partial \theta}\partial_\phi\nonumber\\
&=&t\left[\frac{\delta M}{r_0N_0^2}\partial_t+\frac{2\delta J}{r_0^3}\partial_\phi\right]\theta(r_0-r)\,,
\end{eqnarray}
and can be written as
\beq
h_{\alpha\beta}^{\rm gauge}=\tilde h_{\alpha\beta}^{\rm gauge}\theta(r_0-r)
+\,\mbox{delta-singular terms}\,,
\eeq
with nonvanishing components
\beq
\tilde h^{\rm gauge}_{tt}=\frac{2\delta M}{r}\frac{rN^2}{r_0N_0^2}
\,,\quad
\tilde h^{\rm gauge}_{t\phi}=-\frac{2\delta J}{r}\frac{r^3}{r_0^3}\sin^2\theta\,.
\eeq
Taking into account these gauge modes in the computation of orbital invariants is crucial for obtaining the correct result. 

The final form of the perturbation is then 
\begin{eqnarray}
\label{hnonradschw}
h_{\alpha\beta}^{\rm nonrad}&=&\tilde h_{\alpha\beta}^{\rm gauge}\theta(r_0-r)+\tilde h_{\alpha\beta}^{Z}\theta(r-r_0)\nonumber\\
&&
+\,\mbox{delta-singular terms}\,.
\end{eqnarray}
The completion piece $h_{\alpha\beta}^{\rm comp}$ satisfies the perturbed Einstein's equations \eqref{EFEnonradschw} sourced by the completion part ($l=0,1$) of the particle's energy momentum tensor.
The gauge part $h_{\alpha\beta}^{\rm gauge}$ introduces additional (higher) delta-singular terms (containing also derivatives of the Dirac-delta function at $r=r_0$)\\
\beq
T_{\mu\nu}^{\rm gauge}\equiv \frac{1}{8\pi}\delta G_{\mu\nu}[h_{\alpha\beta}^{\rm gauge}]\,,
\eeq
so that the nonradiative metric perturbation \eqref{hnonradschw} satisfies the field equations
\beq
\delta G_{\mu\nu}[h_{\alpha\beta}^{\rm nonrad}]=8\pi (T_{\mu\nu}^{\rm comp}+T_{\mu\nu}^{\rm gauge})\,.
\eeq
Finally, the Einstein's equations associated with the full perturbation read
\beq
\delta G_{\mu\nu}[h_{\alpha\beta}]=8\pi(T_{\mu\nu}+T_{\mu\nu}^{\rm gauge})\,.
\eeq

We have recalled how the completion problem of metric perturbations is handled in the case of Schwarzschild spacetime and circular orbits.
The original approach of Zerilli suffices in determining the completion part of the nonradiative metric for arbitrary motion, i.e., the exterior metric describing the sudden shift in the spacetime's mass and angular momentum induced by the particle.
Performing a discontinuous gauge transformation then leads to a nonvanishing interior metric, which can be adjusted in order that the whole nonradiative perturbation shares some additional regularity properties, e.g., asymptotic flatness and continuity of certain metric components.
Passing then to the Kerr case this problem is more difficult to be addressed, due to the lack of field equations governing the nonradiative part of the perturbations. 
However, the above considerations apply also to Kerr perturbations, so that we can follow the same line of reasoning as before.

\section{Perturbations on a Kerr spacetime}

Let us consider the Kerr spacetime (with mass $M$ and angular momentum $J=Ma$), whose line element written in Boyer-Lindquist coordinates $x^\alpha=(t,r,\theta,\phi)$, with $\alpha = 0,1,2,3$, is given by
\begin{eqnarray}
\label{kerrmet}
ds^2&=&g_{\alpha\beta}^{K}dx^\alpha dx^\beta
\nonumber\\
&=&\left(1-\frac{2Mr}{\Sigma}  \right) dt^2
+\frac{4aMr \sin^2\theta}{\Sigma}dtd\phi
-\frac{\Sigma}{\Delta}dr^2\nonumber\\
&-& \Sigma d\theta^2 -\left( r^2+a^2+\frac{2Mra^2\sin^2\theta}{\Sigma} \right)\sin^2\theta d\phi^2
\,,\nonumber\\
\end{eqnarray}
where   
\beq
\Delta= r^2+a^2-2Mr\,,\qquad 
\Sigma=r^2+a^2\cos^2\theta\,.
\eeq
The outer horizons $r_+$ is located at $r_+=M+\sqrt{M^2-a^2}$.
Note that the signature has been switched from $+2$ to $-2$ with respect to the Schwarzschild case, as customary within the Teukolsky formalism.

A particle with mass $\mu\ll M$ moving along a geodesic orbit with parametric equations $x_p^\alpha=x_p^\alpha(\tau)$, $\tau$ denoting the proper time parameter, and four velocity $U=\frac{dx_p^\alpha}{d\tau}\partial_\alpha$, with $U\cdot U=1$, induces a perturbation $h_{\alpha\beta}$ on the background due to its energy-momentum tensor, which is Dirac-delta singular along its world line
\beq
\label{Tmunudef}
T_{\mu\nu}(x^\alpha)=\mu \int_{-\infty}^{\infty}\frac1{\sqrt{-g^{K}}}\delta^{(4)}(x^\alpha-x^\alpha_p(\tau))U_\mu U_\nu d\tau
\,,
\eeq
where $g^{K}$ is the metric determinant for the background, and $\delta^{(4)}$ is the four dimensional delta function 
\begin{eqnarray}
\delta^{(4)}(x^\alpha-x^\alpha_p(\tau))&=&\delta(t-t(\tau))\delta^{(3)}(x^a-x^a_p(\tau))\nonumber\\
&=&\frac{\delta(\tau-\tau(t))}{U^t}\delta^{(3)}(x^a-x^a_p(t))
\,,
\end{eqnarray}
with ($a=1,2,3$)
\beq
\delta^{(3)}(x^a-x^a_p(t))=\delta(r-r_p(t))\delta(\theta-\theta_p(t))\delta(\phi-\phi_p(t))
\,.
\eeq
A general geodesic motion is governed by the equations (see, e.g., Ref. \cite{Chandrasekhar:1985kt})
\begin{eqnarray}
\label{geo_eqs}
\frac{d t_p}{d \tau}&=& \frac{1}{\Sigma}\left[aB+\frac{(r_p^2+a^2)}{\Delta}P\right]\,,\qquad
\frac{d r_p}{d \tau}=\epsilon_r \frac{1}{\Sigma}\sqrt{R}\,,\nonumber \\
\frac{d \theta_p}{d \tau}&=&\epsilon_\theta \frac{1}{\Sigma}\sqrt{\Theta}\,,\qquad
\frac{d \phi_p}{d \tau}= \frac{1}{\Sigma}\left[\frac{B}{\sin^2\theta_p}+\frac{a}{\Delta}P\right]\,,
\end{eqnarray}
where $\epsilon_r$ and $\epsilon_\theta$ are sign indicators, and
\begin{eqnarray}
\label{geodefs}
P&=& E(r_p^2+a^2)-aL\,,\quad
B= L-aE \sin^2\theta_p\,, \\
R&=& P^2-\Delta (r_p^2+{\mathcal K})\,,\quad
\Theta={\mathcal K}-a^2\cos^2\theta_p-\frac{B^2}{\sin^2\theta_p}\,.\nonumber
\end{eqnarray}
Here $E=U_t$ and $L=-U_\phi$ denote the conserved Killing energy and angular momentum per unit mass, and ${\mathcal K}$ is a separation constant, usually called the Carter constant, which for equatorial plane orbits ($\theta=\frac{\pi}{2}$, $U^\theta=0$) reduces to ${\mathcal K}=(L-aE)^2$.

The perturbed metric is given by
\beq
\label{fullmet}
g_{\alpha\beta}(x^\lambda)=g_{\alpha\beta}^{K}(x^\lambda)+h_{\alpha\beta}(x^\lambda)\,,
\eeq
where the first order perturbation $h_{\alpha\beta}$ should be suitably regularized, being the retarded field divergent at the particle position. This is usually done by subtracting the Detweiler-Whiting singular field {\it mode-by-mode} through an extension of the particle's 4-velocity $U$ to a field in the neighborhood of the world line and the use of suitable regularization parameters. These issues for radiation gauges and generic orbits in a Kerr spacetime are discussed, e.g., in Ref. \cite{Pound:2013faa}.
The regularized field $h^{\rm R}_{\alpha\beta}$ is thus a smooth vacuum perturbation, with the particle moving along a geodesic in the effective metric $g^{\rm R}_{\alpha\beta}=g^{K}_{\alpha\beta}+h^{\rm R}_{\alpha\beta}$.

We will conveniently use a parametrization of the world line in terms of the coordinate time $t$ instead of the proper time $\tau$. 
On each $t=$ constant spacetime slice the instantaneous position of the particle is represented by a single spatial point, associated with coordinates $x^a=x_p^a(t)$. 
Consider then a spacelike 3-volume $V$ defined by $t=$ constant and $r_+<r_1<r<r_2$, for some values $r_1$ and $r_2$ of the radial variable.
Such a hypersurface is pierced by the particle's world line, which is thus instantaneously bounded by the two 2-surfaces $S_1=V_{r=r_1}$ and $S_2=V_{r=r_2}$, i.e., the inner and outer boundaries of $V$.
Let $r_1=r_p(t)-\epsilon\equiv r_p^-(t)$ and $r_2=r_p(t)+\epsilon\equiv r_p^+(t)$, with $\epsilon\ll1$.
In the limit $\epsilon\to0$ the two spherical-like 2-surfaces $S_1=V_{r=r_p^-(t)}\equiv S_-$ and $S_2=V_{r=r_p^+(t)}\equiv S_+$ join smoothly and are identified with the interface $S=V_{r=r_p(t)}$ between the interior $(-)$ and exterior $(+)$ vacuum regions.
The perturbation can then be written in the form
\beq
h_{\alpha\beta}=h_{\alpha\beta}^{-}\theta(r_p(t)-r)+h_{\alpha\beta}^{+}\theta(r-r_p(t))\,,
\eeq
where $h_{\alpha\beta}^{\pm}$ is given by Eq. \eqref{hdef} and $\theta(x)$ denotes the Heaviside step function.
The regularization procedure basically involves the radiative part $h_{\alpha\beta}^{{\rm rec}\pm}$ of the perturbed metric, whose large-$l$ behavior actually determines the regularization parameters and is enough to obtain a convergent series, once the average of the interior and exterior solutions is taken.

The radiative part of the perturbation is constructed by using the full energy-momentum tensor \eqref{Tmunudef} as a source. 
In fact, according to the CCK procedure in a radiation gauge, the components of the metric perturbation tensor are obtained by applying a suitable differential operator on a scalar quantity, the Hertz potential.
The latter is determined in two steps. It must satisfy the homogeneous Teukolsky equation with opposite spin as the Weyl scalar from which it is constructed (either $\psi_0$ ($s=-2$) or $\psi_4$ ($s=2$)). Furthermore, it must satisfy certain inhomogeneous differential equations sourced by either $\psi_0$ or $\psi_4$, which are in turn the solutions of the inhomogeneous Teukolsky equation sourced by the particle's energy-momentum tensor.
The CCK reconstruction of the metric perturbation from Weyl scalars gives a perturbed metric for which there is no change in the mass and angular momentum, as discussed in Ref. \cite{Keidl:2010pm}. In fact, it is shown there that the retarded metric perturbation constructed from the Hertz potential ($l\geq2$) cannot have the required falloff for large $r$ corresponding to either mass ($l=0$) and angular momentum ($l=1$) perturbations, even if the differential operator acting on the Hertz potential mixes the values of $l$. This result has been confirmed in Refs. \cite{Merlin:2016boc,vandeMeent:2017fqk}, where the mass and angular momentum content of the reconstructed piece of the metric perturbation has been proved to be vanishing either inside or outside of the hypersurface $r=r_p(t)$ containing the particle's world line. This follows from the conservation of the full energy-momentum tensor, which implies that the current $j_\mu\equiv T_{\mu\nu}\xi^\nu$ associated with either the temporal and azimuthal Killing vectors $\xi_{(t)}=\partial_t$ and $\xi_{(\phi)}=\partial_\phi$ of the background spacetime is divergence-free, leading to two Komar-type \lq\lq quasilocally'' conserved integrals across the particle's position \cite{Dolan:2012jg} corresponding to the mass-energy content and the angular-momentum content of the metric perturbation $h_{\alpha\beta}$, respectively. 
These integrals are constructed from the antisymmetric 2-form (see Eq. (121) in Ref. \cite{Merlin:2016boc})
\beq
F_{\mu\nu}[h_{\alpha\beta}]=-\frac1{8\pi}(\xi^\lambda \bar h_{\lambda[\mu;\nu]}+\xi^\lambda{}_{;[\mu}\bar h_{\nu]\lambda}+\xi_{[\mu}\bar h_{\nu]\lambda}{}^{;\lambda})\,,
\eeq 
where $\bar h_{\mu\nu}=h_{\mu\nu}-\frac12g_{\mu\nu}^Kh_\lambda{}^\lambda$ is the trace-reversed metric perturbation, with the property $F_{\mu\nu}{}^{;\nu}=j_\mu$.
Integrating the latter equation over the spacelike 3-volume $V$ introduced above with boundary $\partial V\equiv S_{-}\cup S_{+}$ containing the particle's world line leads to the conserved quantity 
\beq
\label{consquant}
{\mathcal F}(h_{\alpha\beta};\xi,\partial V)=\frac12\int_{\partial V}F^{\mu\nu}dS_{\mu\nu}\,,
\eeq
using Stokes's theorem, where $dS_{\mu\nu}$ denotes the appropriate 2-surface element on $\partial V$.
Computing the above integral associated with either the background temporal and azimuthal Killing vectors then gives the mass and angular momentum content of the perturbation, respectively.
For the radiative part one finds (see Eq. (138) in Ref. \cite{Merlin:2016boc})
\beq
\label{calFhrad}
{\mathcal F}(h_{\alpha\beta}^{\rm rad};\xi_{(t)},\partial V)=0={\mathcal F}(h_{\alpha\beta}^{\rm rad};\xi_{(\phi)},\partial V)\,.
\eeq

A theorem by Wald \cite{Wald:1973} states that a perturbed vacuum metric is determined up to gauge transformations and the addition of perturbations of
the background metric within the family of Petrov type-D vacuum spacetimes, i.e., stationary and axisymmetric perturbations still representing a Kerr black hole but with different mass and angular momentum, or even perturbations of the Kerr metric into other solutions, namely C-metric and Kerr-Newman-Unti-Tamburino (Kerr-NUT) solutions. Therefore, ruling out the latter perturbations for regularity reasons, the remaining task is to determine the changes in mass $\delta M$ and angular momentum $\delta J$ of the background spacetime, i.e., the completion piece, as well as the gauge piece, which represent the nonradiative part of the perturbation
\beq
h_{\alpha\beta}^{\rm nonrad}=h_{\alpha\beta}^{\rm comp}+h_{\alpha\beta}^{\rm gauge}\,.
\eeq  
The completion amplitudes $\delta M$ and $\delta J$ have been determined in Refs. \cite{Merlin:2016boc,vandeMeent:2017fqk} by computing the conserved quantities \eqref{consquant} associated with the completion metric perturbation.
Furthermore, it has been shown there that they are identically zero in the region inside the particle's position, leading to a vanishing interior metric, while in the exterior region they are given by the conserved energy and angular momentum of the particle.
The completion metric thus results in a sudden shift in mass and angular momentum parameters induced by the particle occurring at $r=r_p(t)$, just as in the Schwarzschild case in the Zerilli gauge (see Eq. \eqref{hcompschw}).

The results of Refs. \cite{Merlin:2016boc,vandeMeent:2017fqk} have been obtained under the assumption of bound motion, though they are likely to hold also for arbitrary trajectories. 
Therefore, we will limit our considerations to bound orbits too.

\subsection{Completion piece}

The exterior metric $h_{\alpha\beta}^{{\rm comp}+}$ is obtained by first-order mass and angular momentum perturbations of the Kerr solution (so that it is stationary and axisymmetric too), i.e., 
\beq
h_{\alpha\beta}^{{\rm comp}+}=h^{{\rm comp}+, \delta M}_{\alpha\beta}\delta M+h^{{\rm comp}+, \delta J}_{\alpha\beta}\delta J\,,
\eeq
where the change in mass $h^{{\rm comp}+, \delta M}_{\alpha\beta}=\partial_M g^{K}_{\alpha\beta}$ and angular momentum $h^{{\rm comp}+, \delta J}_{\alpha\beta}=\partial_J g^{K}_{\alpha\beta}$ are obtained by replacing $M\to M+\delta M$ and $J\to J+\delta J$ in the background metric \eqref{kerrmet} and retaining only terms which are linear in $\delta M$ and $\delta J$ \cite{Merlin:2016boc}. The latter are then identified with the conserved energy and angular momentum of the perturbing particle, i.e., $\delta M=\mu E$ and $\delta J=\mu L$.
We list below the nonvanishing components of $h_{\alpha\beta}^{{\rm comp}+}$, for completeness (see Eqs. (88)--(89) of Ref. \cite{Merlin:2016boc})
\begin{eqnarray}
h^{{\rm comp}+}_{tt}&=&
-\frac{2r}{\Sigma^2}[(r^2+3a^2\cos^2\theta)\delta M-2a\cos^2\theta\delta J]
\,,\nonumber\\
h^{{\rm comp}+}_{rr}&=&
-\frac{2r}{M\Delta^2}\{[M(r^2+3a^2\cos^2\theta)+a^2r\sin^2\theta]\delta M \nonumber\\
&&-a[r-(r-2M)\cos^2\theta]\delta J\}
\,,\nonumber\\ 
h^{{\rm comp}+}_{\theta\theta}&=&
\frac{2a}{M}\cos^2\theta(a\delta M-\delta J)
\,,\nonumber\\ 
h^{{\rm comp}+}_{\phi\phi}&=&
\frac{2a}{M\Sigma^2}\sin^2\theta\{a[\Sigma^2
+Mr\sin^2\theta(r^2\nonumber\\
&&
-a^2\cos^2\theta)]\delta M
-(\Sigma^2+2Mr^3\sin^2\theta)\delta J\}
\,,\nonumber\\
h^{{\rm comp}+}_{t\phi}&=&
\frac{2r}{\Sigma^2}\sin^2\theta[2a^3\cos^2\theta\delta M
+(r^2-a^2\cos^2\theta)\delta J]
\,.\nonumber\\
\end{eqnarray}
The interior metric $h_{\alpha\beta}^{{\rm comp}-}$ is instead identically vanishing, as already stated, so that 
\beq
h_{\alpha\beta}^{\rm comp}=h_{\alpha\beta}^{{\rm comp}+}\theta(r-r_p(t))\,.
\eeq
Its double contraction with the particle's four velocity is then given by
\beq
h_{UU}^{\rm comp}=h_{\alpha\beta}^{\rm comp}U^\alpha U^\beta
=h_{UU}^{{\rm comp}+}\theta(r-r_p(t))\,.
\eeq 

The completion metric perturbation is thus associated with the energy-momentum tensor 
\beq
T_{\mu\nu}^{\rm comp}\equiv\frac1{8\pi}\delta G_{\mu\nu}[h_{\alpha\beta}^{\rm comp}]\,,
\eeq
the components of which are proportional to the Dirac-delta function and its first derivatives, i.e., 
\beq
T_{\mu\nu}^{\rm comp}=\tilde T_{\mu\nu}^{\rm comp,\delta}\delta(r-r_p(t))+\tilde T_{\mu\nu}^{\rm comp,\delta'}\delta'(r-r_p(t))\,.
\eeq

\subsection{Gauge piece}

The gauge part $h_{\alpha\beta}^{{\rm gauge}\pm}$ of the perturbed metric is given by 
\beq
h_{\alpha\beta}^{{\rm gauge}}=2\xi_{(\alpha; \beta)}\,,
\eeq
with a discontinuous gauge field
\beq
\xi=\xi_-\theta(r_p(t)-r)+\xi_+\theta(r-r_p(t))\,,
\eeq
which generates delta-singular terms in the metric and consequently higher and higher singular terms in the connection, the curvature, etc.
As in the Schwarzschild case, one can equivalently write
\begin{eqnarray}
h_{\alpha\beta}^{{\rm gauge}}&=&h_{\alpha\beta}^{{\rm gauge}-}\theta(r_p(t)-r)+h_{\alpha\beta}^{{\rm gauge}+}\theta(r-r_p(t))\nonumber\\
&&
+\,\mbox{delta-singular terms}\,,
\end{eqnarray}
where
\beq
h_{\alpha\beta}^{{\rm gauge}\pm}=2\xi_{\pm\,(\alpha; \beta)}\,.
\eeq
The interior and exterior parts $\xi_\pm$ of the gauge field have the general form \cite{Shah:2015nva,Bini:2018ylh} 
\beq
\xi_\pm=\frac{\mu}{M}\left[\alpha_\pm(t)\partial_t +\beta_\pm(t)\partial_\phi\right]\,.
\eeq
The only nonvanishing components of $h^{{\rm gauge}\pm}_{\alpha\beta}$ then turn out to be
\begin{eqnarray}
h^{{\rm gauge}\pm}_{tt}&=&
2\frac{\mu}{M}\left[\frac{d\alpha_\pm}{dt}\left(1-\frac{2Mr}{\Sigma}\right)\right.\nonumber\\
&&\left.
+\frac{d\beta_\pm}{dt}\frac{2aMr}{\Sigma}\sin^2\theta\right]
\,,\nonumber\\
h^{{\rm gauge}\pm}_{t\phi}&=&
\frac{\mu}{M}\left\{\frac{d\alpha_\pm}{dt}\frac{2Mar}{\Sigma}\right.\nonumber\\
&&\left.
-\frac{d\beta_\pm}{dt}\left[\Delta +\frac{2Mr}{\Sigma}(r^2+a^2)\right]\right\}\sin^2\theta
\,,\nonumber\\
\end{eqnarray}
which we require bounded in time and asymptotically flat (i.e., they must vanish for large $r$).
The latter request implies that $h^{{\rm gauge}+}_{\alpha\beta}\equiv0$, so that $\alpha_+(t)=0=\beta_+(t)$, leading to 
\begin{eqnarray}
\label{eq_h_gauge}
h_{\alpha\beta}^{{\rm gauge}}&=&h_{\alpha\beta}^{{\rm gauge}-}\theta(r_p(t)-r)
+h_{\alpha\beta}^{\rm gauge,\delta}\delta(r-r_p(t))\,.\nonumber\\
\end{eqnarray}
Only the functions $\alpha_-(t)$ and $\beta_-(t)$ remain to be specified.
Therefore, we need two further conditions.

The singular part of the gauge perturbation (according to the notation of Eq. \eqref{eq_h_gauge}) has nonvanishing components
\begin{eqnarray}
h^{\rm gauge,\delta}_{tt}
&=&-2\frac{dr_p}{dt}h^{\rm gauge,\delta}_{tr}
\,,\nonumber\\
h^{\rm gauge,\delta}_{tr}&=&
-\frac{\mu}{M}\left[\alpha_-\left(1-\frac{2Mr_p}{\Sigma}\right)
+\beta_-\frac{2aMr_p}{\Sigma}\sin^2\theta\right]
\,,\nonumber\\
h^{\rm gauge,\delta}_{t\phi}
&=&-\frac{dr_p}{dt}h^{\rm gauge,\delta}_{r\phi}
\,,\nonumber\\
h^{\rm gauge,\delta}_{r\phi}&=&
-\frac{\mu}{M}\left\{\alpha_-\frac{2Mar_p}{\Sigma}\right.\nonumber\\
&&\left.
-\beta_-\left[\Delta +\frac{2Mr_p}{\Sigma}(r_p^2+a^2)\right]\right\}\sin^2\theta
\,,
\end{eqnarray}
so that the double contraction with the particle's four velocity is identically vanishing (i.e.,  $h_{UU}^{\rm gauge,\delta}=0$), whence
\beq
h_{UU}^{\rm gauge}
=h_{UU}^{\rm gauge-}\theta(r_p(t)-r)\,.
\eeq 

Adding gauge modes to the completion piece thus generates a further high singular source term 
\beq
T_{\mu\nu}^{\rm gauge}\equiv\frac{1}{8\pi}\delta G_{\mu\nu}[h_{\alpha\beta}^{\rm gauge}]\,,
\eeq
with
\begin{eqnarray}
T_{\mu\nu}^{\rm gauge}&=&\tilde T_{\mu\nu}^{\rm gauge,\delta}\delta(r-r_p(t))+\tilde T_{\mu\nu}^{\rm gauge,\delta'}\delta'(r-r_p(t))\nonumber\\
&&
+\tilde T_{\mu\nu}^{\rm gauge,\delta''}\delta''(r-r_p(t))\,.
\end{eqnarray}

\subsection{Gauge-fixing}

The resulting metric perturbation associated with the nonradiative multipoles is then
\begin{eqnarray}
\label{met01}
h^{\rm nonrad}_{\alpha\beta}&=&h_{\alpha\beta}^{{\rm gauge}-}\theta(r_p(t)-r)+h_{\alpha\beta}^{{\rm comp}+}\theta(r-r_p(t))\nonumber\\
&&
+h_{\alpha\beta}^{\rm gauge,\delta}\delta(r-r_p(t))\,.
\end{eqnarray}
Each metric component is a smooth function of the coordinates in either side, and is well behaved in the limit $r\to r_p(t)$.  
However, the regular part of the metric is not continuous at the particle position, and derivatives generate distributional singularities there, which add to those arising from the singular part of the metric.
Therefore, all relevant tensors associated with the metric \eqref{met01} are meant in the sense of distributions.

The first condition comes from demanding that the causality property of the particle's four velocity be preserved with respect to the full (regularized) perturbed metric $g^{\rm R}_{\alpha\beta}$ at every spacetime point, including the location of the particle.
This is equivalent to impose the continuity of $h^{\rm R}_{UU}=h^{\rm R}_{\alpha\beta}U^\alpha U^\beta$ across the instantaneous position of the particle $r=r_p(t)$, which for the metric \eqref{met01} implies
\beq
\label{h_uu_cond}
h_{UU}^{{\rm comp}+}\vert_{x^\alpha=x^\alpha_p(t)}=h_{UU}^{{\rm gauge}-}\vert_{x^\alpha=x^\alpha_p(t)}\,.
\eeq

The second condition can be derived as follows.
Consider the full spacetime metric
\beq
g_{\alpha\beta}=g_{\alpha\beta}^{K}+h^{\rm rec}_{\alpha\beta}+h^{\rm nonrad}_{\alpha\beta}\,,
\eeq 
satisfying the Einstein's equations 
\beq
G_{\mu\nu}[g_{\alpha\beta}]=8\pi T_{\mu\nu}\,.
\eeq
For an arbitrary vector field $v=v^\alpha \partial_\alpha$ the exterior derivative writes
\beq
dv=\frac12(dv)_{\alpha\beta} dx^\alpha \wedge dx^\beta
\,,
\eeq
with components 
\beq
\label{extdiff}
(dv)_{\alpha\beta}=2v_{[\beta;\alpha]}\,.
\eeq
The Ricci identity and the Einstein equations imply \cite{Chandrasekhar:1985kt}
\begin{eqnarray}
\label{ricciid}
v^\alpha{}_{;\beta\alpha}-v^\alpha{}_{;\alpha\beta}
&=&R^\alpha{}_{\gamma\alpha\beta}v^\gamma\nonumber\\
&=&R_{\gamma\beta}v^\gamma
=-8\pi T^{\rm(TR)}_{\gamma\beta}v^\gamma\,,
\end{eqnarray}
where we have introduced the trace-reversed (TR) notation
\beq
T^{\rm (TR)}_{\alpha\beta}=T_{\alpha\beta} -\frac12Tg_{\alpha\beta}\,,\qquad 
T=T^\alpha{}_\alpha\,.
\eeq
The energy-momentum tensor is proportional to the mass ratio $\mu/M$ like the components of the perturbed metric, so that indices can be raised/lowered with the background metric to first order.
From Eq. \eqref{Tmunudef} we then have for the particle's energy-momentum tensor
\begin{eqnarray}
\label{Tmunutr}
T^{\rm (TR)}_{\alpha\beta}&=&T\left(U_\alpha U_{\beta}-\frac12 g^K_{\alpha\beta}\right)\,,\nonumber\\
T&=& \frac{\mu}{\Sigma_p \sin\theta_p U^t}\delta^{(3)}(x-x_p(t))\,.
\end{eqnarray}

Taking the covariant derivative of both sides of Eq. \eqref{extdiff} then gives 
\beq
\label{cddv}
(dv)_{\alpha\beta}{}^{;\beta}=-8\pi\, T^{\rm (TR)}_{\alpha\beta} \, v^\beta
+v^\beta{}_{;\beta\alpha}-v_{\alpha;\beta}{}^{\beta}\,.
\eeq
The last term can be conveniently rewritten by separating its symmetric and antisymmetric parts as follows
\beq
v_{\alpha;\beta}{}^{\beta}=v_{[\alpha;\beta]}{}^{\beta}+v_{(\alpha;\beta)}{}^{\beta}
=-\frac12(dv)_{\alpha\beta}{}^{;\beta}+v_{(\alpha;\beta)}{}^{\beta}\,,
\eeq
whence Eq. \eqref{cddv} becomes
\beq
\label{cddv2}
\frac12(dv)_{\alpha\beta}{}^{;\beta}=-8\pi T^{\rm (TR)}_{\alpha\beta} v^\beta
+v^\beta{}_{;\beta\alpha}-v_{(\alpha;\beta)}{}^{\beta}\,.
\eeq
Note that if $v$ is a Killing vector of the perturbed spacetime, i.e., $v=K$, the last two terms vanish identically (because of the relations $K^\beta{}_{;\beta}=0=K_{(\alpha;\beta)}$) and Eq. \eqref{cddv2} simplifies to 
\beq
(dK)_{\alpha\beta}{}^{;\beta}=-16 \pi\, T^{\rm (TR)}_{\alpha\beta} K^\beta\,.
\eeq

Let us now integrate both sides of Eq. \eqref{cddv2} over the spacelike 3-volume $V$ with boundary $S_{-}\cup S_{+}$ introduced above, surrounding the instantaneous location of the particle $r=r_p(t)$.
This leads to the relation
\beq
\label{cddv2int}
\frac12{\mathcal I}_1={\mathcal I}_2+{\mathcal I}_3\,,
\eeq
where
\begin{eqnarray}
\label{Idefs}
{\mathcal I}_1&\equiv&\int_V (dv)_{\alpha\beta}{}^{;\beta}dV^\alpha\nonumber\\
&=&
\int_{S_{+}} (dv)_{\alpha\beta}dS^{\alpha\beta}-\int_{S_{-}} (dv)_{\alpha\beta}dS^{\alpha\beta}
\,,\nonumber\\
{\mathcal I}_2&\equiv&
-8\pi\int_V T^{\rm (TR)}_{\alpha\beta} \, v^\beta dV^\alpha
\,,\nonumber\\
{\mathcal I}_3&\equiv&
\int_V [v^\beta{}_{;\beta\alpha}-v_{(\alpha;\beta)}{}^{\beta}]dV^\alpha
\,,
\end{eqnarray}
which allows one to connect interior and exterior metrics at the boundary.
The volume element $dV^\alpha$ and the surface element $dS^{\alpha\beta}$ are defined by
\beq
dV^\alpha=\eta^\alpha{}_{r\theta\phi} dr d\theta d\phi\,,\qquad
dS^{\alpha\beta}=\eta^{\alpha\beta}{}_{\theta\phi} d\theta d\phi\,,
\eeq
respectively, where $\eta_{\alpha\beta\gamma\delta}=\sqrt{-g}\, \epsilon_{\alpha\beta\gamma\delta}$ is the unit volume 4-form, with $\epsilon_{\alpha\beta\gamma\delta}$ ($\epsilon_{0123}=1$) being the Levi-Civita alternating symbol.
Higher singular terms containing derivatives of the Dirac-delta function do not contribute to the integrals \eqref{Idefs}. 

The relations derived above are completely general, and are valid for any choice of the vector field $v$ (with corresponding 1-form $v^\flat$).
The latter can be naturally chosen as aligned with either the temporal or azimuthal Killing vector of the background spacetime, or even a combination of them, like the four velocity of the zero-angular-momentum observers (ZAMOs).
For the application we are going to discuss below we will take
\begin{eqnarray}
v&=&\partial_t\,,\nonumber\\
v^{\flat\pm}&=&g_{t\alpha}dx^\alpha=(g^{K}_{t\alpha}+h^{\pm}_{t\alpha})dx^\alpha\,,
\end{eqnarray}
which implies different expressions for the exterior derivatives $dv^\pm$ in either region.
In contrast, when considered as a vector and not as a 1-form, its components are continuous across the particle's location.
Notice that choosing $v=\partial_\phi$ would not give any useful information.
In fact, it is a Killing vector for both interior and exterior spacetime regions, implying that ${\mathcal I}_3\equiv0$ and the relation \eqref{cddv2int} reduces to a trivial identity not involving the functions $\alpha_-(t)$ and $\beta_-(t)$.

We have recalled above how the full particle's energy momentum tensor $T_{\mu\nu}$ fixes the completion amplitude by constructing with it Komar-like integrals over a closed surface enveloping the region containing the matter distribution, thus completely determining the completion piece.
The radiative part of the perturbation does not carry any mass and angular momentum, as it follows from computing the conserved quantities associated with the currents $j_\mu=T_{\mu\nu}\xi^\nu$ built with the background Killing vectors $\xi_{(t)}=\partial_t$ and $\xi_{(\phi)}=\partial_\phi$ (see Eq. \eqref{calFhrad}).
Applying then this result to the integral form \eqref{cddv2int}--\eqref{Idefs} of the Ricci identity \eqref{ricciid}, where $v=\partial_t$ is a Killing vector for the background spacetime, we have that the radiative part of the metric perturbation does not contribute to the integrals ${\mathcal I}_1$ and ${\mathcal I}_3$.
Hence, the latter can be computed simply by using the nonradiative part of the metric.

Let us consider the case of eccentric equatorial orbits, as an example.
The particle's four velocity is given by
\begin{eqnarray}
\label{barudef}
U&=&\frac{1}{r_p^2} \left(a x+\frac{r_p^2+a^2}{\Delta} P\right)\partial_t\nonumber\\
&+&
\frac{\epsilon_r}{r_p^2}\left[P^2-\Delta(r_p^2+x^2)\right]^{1/2}\partial_r 
+\frac{1}{r_p^2}\left(x+\frac{a}{\Delta} P\right)\partial_\phi
\,,\nonumber\\
\end{eqnarray}
as from Eq. \eqref{geodefs}, with $x=L-aE$.
The first two integrals \eqref{Idefs} are easily computed
\begin{eqnarray}
{\mathcal I}_1&=& -8\pi \left[\delta M -\mu\left(\frac{d\alpha_-}{dt}-2a\frac{d\beta_-}{dt}\right)\right]
\,,\nonumber\\
{\mathcal I}_2&=&-8\pi\mu\left(E-\frac1{2U^t} \right)
\,.
\end{eqnarray}
The evaluation of the third integral ${\mathcal I}_3$ instead is much more involved, since differentiating twice the Heaviside function generates terms proportional to the Dirac-delta function and its first derivatives.
We find
\beq
{\mathcal I}_3={\mathcal I}_3^{\delta M}\delta M+{\mathcal I}_3^{\delta J}\delta J\,,
\eeq
with
\begin{widetext}
\begin{eqnarray}
{\mathcal I}_3^{\delta M}&=&-a{\mathcal I}_3^{\delta J}
+4\pi r_p\frac{(r_p^2+a^2)^2}{\Delta^2}\left[
\frac{d^2r_p}{dt^2}+\frac1{r_p}\left(\frac{dr_p}{dt}\right)^2\left(
1-\frac{4Mr_p}{\Delta}\frac{r_p^2-a^2}{r_p^2+a^2}
\right)\right]
\,,\\
{\mathcal I}_3^{\delta J}&=&
-\frac{8\pi r_p}{a}\frac{r_p^2+a^2}{\Delta}\left\{
\frac{d^2r_p}{dt^2}\left[\frac{r_p}{a}\,{\rm arctan}\left(\frac{a}{r_p}\right)
-1+\frac{a^2}{\Delta}-\frac{a^2}{3Mr_p}\left(1+\frac{Mr_p}{r_p^2+a^2}\right)
\right]\right.\nonumber\\
&&\left.
+\frac2{r_p}\left(\frac{dr_p}{dt}\right)^2\left(
1-\frac{Mr_p}{\Delta}\frac{r_p^2-a^2}{r_p^2+a^2}
\right)
\left[\frac{r_p}{a}\,{\rm arctan}\left(\frac{a}{r_p}\right)
-1+\frac{2a^2}{\Delta}-\frac12\frac{a^2[2Mr_p+3(r_p^2+a^2)]}{(r_p^2+a^2)^2-Mr_p(3r_p^2+a^2)}
\right]\right\}
\,.\nonumber
\end{eqnarray}
The continuity \eqref{h_uu_cond} of $h^{\rm nonrad}_{UU}$ at the particle's position $r=r_p(t)$ reads
\begin{eqnarray}
\label{h_uu_cond_ecc}
h_{UU}^{{\rm comp}+}(r_p)&=&
-2\frac{\delta M}{r_p}(U^t)^2
+4\frac{\delta J}{r_p}U^tU^{\phi}
-2\frac{r_p^2}{M\Delta^2}[(Mr_p+a^2)\delta M-a\delta J](U^r)^2\nonumber\\
&&
+2\frac{a}{M}\left[a\left(1+\frac{M}{r_p}\right)\delta M-\left(1+\frac{2M}{r_p}\right)\delta J\right](U^{\phi})^2\nonumber\\
&=&2\frac{\mu}{M}U^t\left(E\frac{d\alpha_-}{dt}-L\frac{d\beta_-}{dt}\right)
=h_{UU}^{{\rm gauge}-}(r_p)
\,.
\end{eqnarray}
The two conditions \eqref{cddv2int} and \eqref{h_uu_cond_ecc} can be finally solved for the time derivatives of $\alpha_-$ and $\beta_-$, leading to
\begin{eqnarray}
\label{solabfin}
(L-2aE)\frac{d\alpha_-}{dt}&=&
EL+\frac{L}{4\pi\mu}({\mathcal I}_2+{\mathcal I}_3)-\frac{Ma}{\mu U^t}h_{UU}^{{\rm comp}+}(r_p)
\,,\nonumber\\
(L-2aE)\frac{d\beta_-}{dt}&=&
E^2+\frac{E}{4\pi\mu}({\mathcal I}_2+{\mathcal I}_3)-\frac{M}{2\mu U^t}h_{UU}^{{\rm comp}+}(r_p)
\,,
\end{eqnarray}
\end{widetext}
where the quantities ${\mathcal I}_2$, ${\mathcal I}_3$ and $h_{UU}^{{\rm comp}+}(r_p)$ are all functions of $r_p$ only, with $\delta M=\mu E$ and $\delta J=\mu L$ as already stated.

This result has been successfully applied in Ref. \cite{Bini:2019lkm} to compute the first-order GSF correction to the gyroscope precession along slightly eccentric orbits and the related periastron advance in the circular orbit limit \cite{Bini:2019zjj}.
In the circular case ($r=r_0$, $U^r=0$) the previous relations reduce to
\begin{eqnarray}
\frac{d\alpha_-}{dt}&=&
-\frac{u_0(1+2\hat au_0^{3/2}-\hat a^2u_0^2)}{(1-3u_0+2\hat au_0^{3/2})^{1/2}(1+\hat au_0^{3/2})}
\,,\nonumber\\
M\frac{d\beta_-}{dt}&=&
-\frac{u_0^{5/2}(2-\hat au_0^{1/2})}{(1-3u_0+2\hat au_0^{3/2})^{1/2}(1+\hat au_0^{3/2})}
\,,
\end{eqnarray}
where $u_0=M/r_0$ and $\hat a=a/M$, so that the corresponding gauge vector coincides with that used in Ref. \cite{Bini:2018ylh} (see Eqs. (3.14)--(3.16) there).

In the Schwarzschild limit ($a=0$) Eq. \eqref{solabfin} becomes
\begin{eqnarray}
\frac{d\alpha_-}{dt}&=&
\frac{ML^2}{r_p^3E} +\frac{M}{r_pE}-\frac{2ME}{r_p-2 M}
\,,\nonumber\\
M\frac{d\beta_-}{dt}&=&
-\frac{2M^2L}{r_p^3}
\,,
\end{eqnarray}
which for circular orbits reduces to 
\beq
\frac{d\alpha_-}{dt}=-\frac{u_0}{\sqrt{1-3u_0}}\,,\qquad
M\frac{d\beta_-}{dt}=-\frac{2u_0^{5/2}}{\sqrt{1-3u_0}}\,,
\eeq
in agreement with  Eq. \eqref{xischwcirc}.
The resulting metric components thus reproduce previous results for both circular and eccentric orbits \cite{Detweiler:2003ci,Bini:2014ica,Hopper:2015icj}.

\section{Discussion}

GSF calculations of orbital invariants require the complete knowledge of the perturbed metric, including a radiative part and a completion piece, made of nonradiative modes and gauge modes. 
The gauge-smoothing of the perturbation on a Kerr spacetime across the particle position has been a challenge for many years.
We have finally solved this problem, providing a prescription for fully determining the gauge part of the metric perturbation, resulting in two conditions to be imposed on the completion piece of the metric. These conditions are necessary to preserve the causality property of the particle's four velocity at its instantaneous position (see Eq. \eqref{h_uu_cond}), and to fulfill the Ricci identity across the hypersurface containing the particle's world line (see Eqs. \eqref{cddv2int}--\eqref{Idefs}).

We have applied this prescription to the case of eccentric equatorial orbits, as an example.
The resulting nonradiative metric perturbation has been used in Refs. \cite{Bini:2019lkm,Bini:2019zjj} to compute the first-order GSF correction to the gyroscope precession along slightly eccentric orbits and the related periastron advance in the circular orbit limit.
We have also recovered all limiting cases already studied in the literature of eccentric orbits in  Schwarzschild as well as circular orbits in both Schwarzschild and Kerr spacetimes. 
This result will allow for completing ongoing and future GSF calculations, including the case of inclined orbits, i.e., bound orbits not confined to the equatorial plane, which are expected to play a key role in the study of orbital resonances in EMRIs, strongly affecting the phasing of the inspiral \cite{Flanagan:2010cd}.

\section*{Acknowledgements}

DB thanks the organizers of the 19th Capra meeting for devoting an open discussion to the problem of the low multipoles in the Kerr spacetime. 
He also thanks L. Barack, T. Damour, C. Kavanagh and M. Van de Meent for helpful comments.
Both authors thank the anonymous referees for providing useful criticisms, corrections and suggestions to improve the manuscript presentation.

\end{document}